\documentclass[preprint,preprintnumbers,amsmath,amssymb]{revtex4}

\usepackage{graphicx}
\usepackage{dcolumn}
\usepackage{bm}

\begin{document}
\newcommand{\kp}{{\bf k$\cdot$p}\ }

\begin{center}
	\textbf{\Large Comment}
\end{center}

The following paper on the band structure of Cd$_3$As$_2$ was published in a camera-ready form by my graduate student  Jan Bodnar in the Proceedings of the 3rd International Conference on the Physics of Narrow Gap Semiconductors, edited by J. Rauluszkiewicz, M. Gorska and E. Kaczmarek,  (PWN-Polish Scientific Publishers,  Warszawa 1977), p. 311-316. A few months after the conference Jan died suddenly of a heart attack at the age of 27. Because lately Cd$_3$As$_2$ has become an important material in the field of topological matter, while the Proceedings of NGS Warsaw Conference are at present hardly available, I felt that Bodnar's work  should become accessible on arXiv. The paper appears without changes,  some misprints have been corrected. I thank  my colleagues  Pawel Pfeffer and Grzegorz Mazur for friendly help with the  preparation of electronic version.
\ \\
Wlodek Zawadzki, [zawad@ifpan.edu.pl]

\title{
Band structure of Cd$_3$As$_2$ from Shubnikov - de Haas

 and de Haas - van Alphen effects }

\author{\framebox{Jan Bodnar}}
 \affiliation{Department of Solid State Physics, Polish Academy of Sciences, Zabrze, Poland}

\begin{abstract}

Experimental values of SdH and dHvA periods and cyclotron effective masses found by Rosenman and Doi et al. have been compared with the theoretical predictions derived in this work for a tetragonal narrow gap semiconductor. By the least square fit method the values of band parameters were obtained. It has been established that Cd$_3$As$_2$ has inverted band structure resembling HgTe under tensile stress.

\end{abstract}

\maketitle

\section{\label{sec:level1}INTRODUCTION\protect\\ \lowercase{}}

In heavily doped semiconductors, such as Cd$_3$As$_2$ (n$_{min}\approx 10^{18}$ cm$^{-3}$ [1]), measurements of SdH and dHvA effects serve as powerful tools in band structure investigations. In this work we analyse SdH and dHvA experiments in Cd$_3$As$_2$ [2-4] using a theory which presents a generalization of the Kane band model [5] for the case of a tetragonal semiconductor with small energy gap [6].

\section{\label{sec:level2}THEORY\protect\\ \lowercase{}}

An influence of the tetragonal field on the three-level Kane band model may be characterized by a parameter $\delta$, which represents the crystal field splitting of p-type levels. In both cases of the simple or inverted level ordering tetragonal field splits the fourfold degenerate state $\Gamma_8$ into twofold degenerate levels, thus giving a four-level scheme. Considering exactly the {\kp} coupling between these four levels we have obtained the following secular equation  for the energies:

\begin{equation}
\gamma({\cal E}) = f_2({\cal E}) \cdot (k^2_x + k^2_y) + f_1({\cal E}) \cdot k^2_z,
\end{equation}
where
\begin{equation}
\gamma({\cal E}) = ({\cal E}-{\cal E}_c)\left({\cal E}-{\cal E}_v-\frac{\Delta}{3}\right)
\left[\left({\cal E}-{\cal E}_v+\frac{2 \Delta}{3}\right)\left({\cal E}-{\cal E}_v-\frac{\Delta}{3}\right)+\delta\left({\cal E}-{\cal E}_v+\frac{\Delta}{3}\right)\right],
\end{equation}
with
\begin{equation}
f_1({\cal E}) = P^2_{||}\left({\cal E}-{\cal E}_v-\frac{\Delta}{3}\right)\left({\cal E}-{\cal E}_v+\frac{\Delta}{3}\right),
\end{equation}
\begin{equation}
f_2({\cal E}) = P^2_{\perp}\left[\left({\cal E}-{\cal E}_v-\frac{\Delta}{3}\right)\left({\cal E}-{\cal E}_v+\frac{\Delta}{3}\right)
+\delta({\cal E}-{\cal E}_v)\right].
\end{equation}
 ${\cal E}$ is the energy of electrons or holes, P$_{\perp}$, P$_{||}$ and $\Delta$ are matrix elements of the momentum operator and of the spin-orbit interaction, respectively. ${\cal E}_c$ and ${\cal E}_v$ denote energies of the band edge states for the Hamiltonian without the spin-orbit coupling. These may be expressed by the energy gap ${\cal E}_g={\cal E}(\Gamma_{6s})-{\cal E}(\Gamma_{7p})$. Choosing suitable values of ${\cal E}_c$ and ${\cal E}_v$ we may describe with the help of Eq.(1) all possible four-band models of the tetragonal semiconductor with a small energy gap, i.e. simple or inverted band structure for $\delta >$0 and $\delta <$0. Equation of the type (1) was first derived by Kildal [7] for CdGeAs$_2$ under additional restrictions. It follows from Eq.(1) that the conduction band has simple ellipsoidal shape in k-space with the anisotropy depending, however, on the energy.
\begin{figure}
\includegraphics[scale=0.35,angle=0, bb = 550 20 240 600]{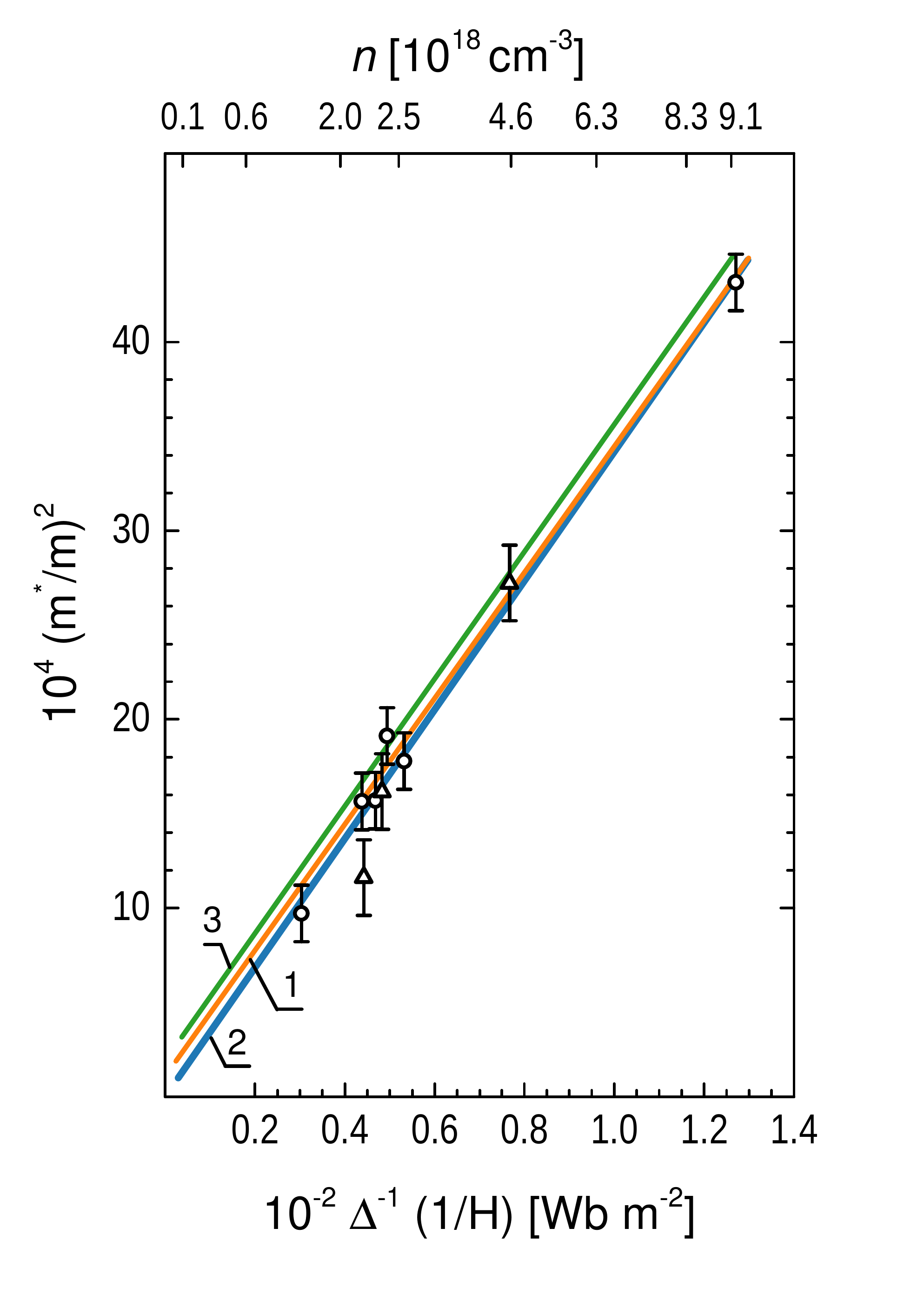}
\caption{\label{fig:epsart}{Dependence of effective mass squared in Cd$_3$As$_2$ on the inverse oscillation period. The corresponding electron concentrations are indicated on the upper abscissa. Solid lines -
theory, circles - SdH data, triangles - dHvA data.
}} \label{fig1th}
\end{figure}
\begin{figure}
\includegraphics[scale=0.35,angle=0, bb = 550 20 240 600]{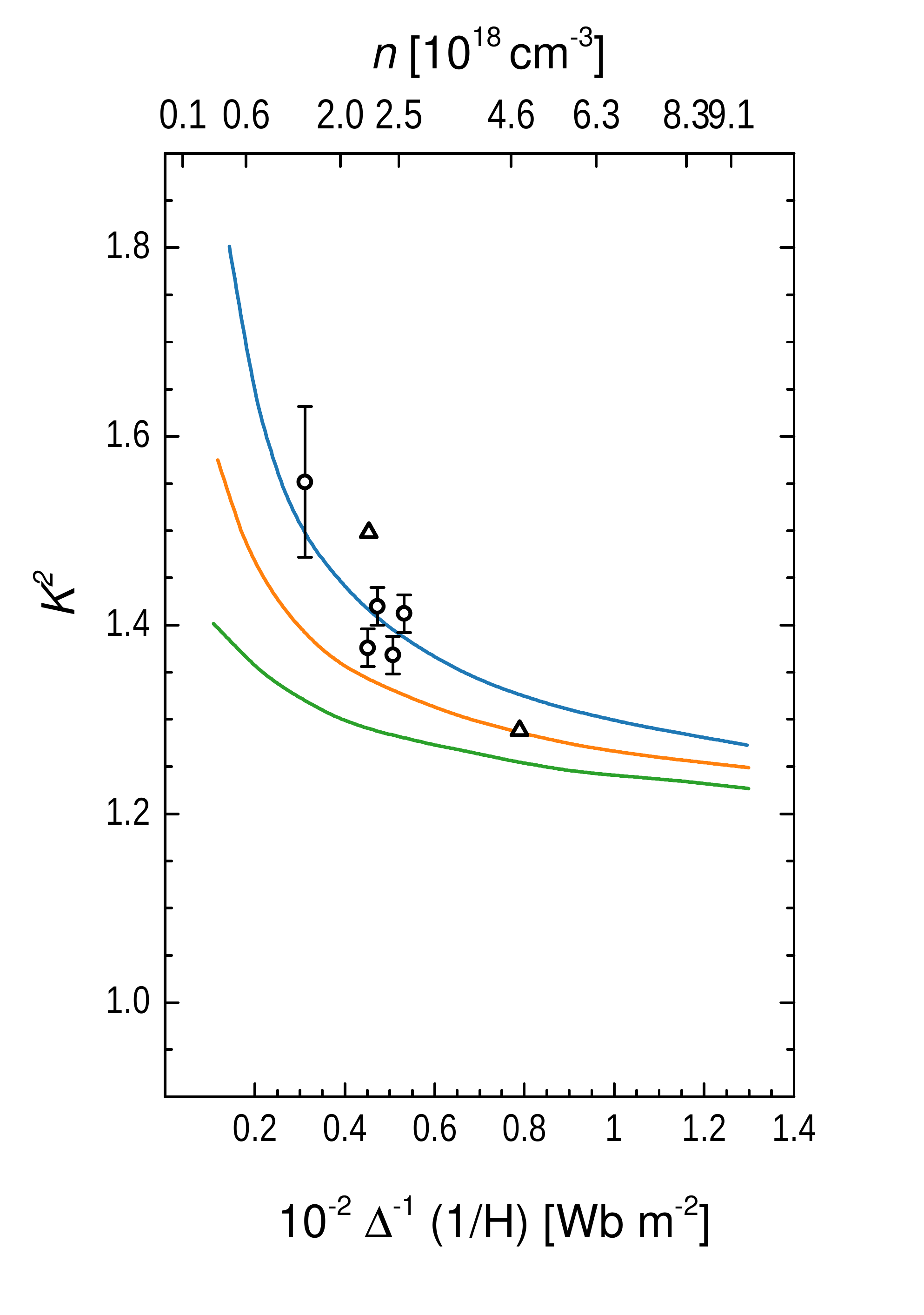}
\caption{\label{fig:epsart}{Dependence of the anisotropy parameter squared in Cd$_3$As$_2$ on the inverse oscillation period. The corresponding electron concentrations are indicated on the upper abscissa. Solid lines -
theory, circles - SdH data, triangles - dHvA data.
}} \label{fig2th}
\end{figure}

Extreme cross-section of the Fermi ellipsoid in the plane perpendicular to the magnetic field \textbf{H} is given by
\begin{equation}
S_m=\pi\gamma({\cal E})[f_2({\cal E})]^{-1/2}[f_2({\cal E})cos^2\Theta + f_1({\cal E})sin^2\Theta]^{-1/2},
\end{equation}
where $\Theta$ is an angle between the magnetic field and the tetragonal axis. According to the semiclassical theory of Lifshitz and Kosevich [8] a period $\Delta(1/H)$ of SdH and dHvA oscillations and the cyclotron effective mass are related to S$_m$ as follows

\begin{equation}
\Delta(1/H)=2\pi e/c\hbar S_m\;,\;\;\;\;\;\;\;\;\;\;m^*({\cal E},\Theta) = (\hbar^2/2\pi)(\delta S_m/\delta{\cal E}).
\end{equation}
From expressions (6) we may determine the coefficient of anisotropy

\begin{equation}
K = \Delta_{||}(\Theta = 0^{\circ})/\Delta_{\perp}(\Theta = 90^{\circ}) = [f_2({\cal E})/f_1({\cal E})]^{1/2}.
\end{equation}
Equations (1)-(6) have been used to compare and fit the theoretical dependence of K and m$^*$ to available experimental data.

\section{\label{sec:level2}Numerical results\protect\\ \lowercase{}}

\begin{figure}
\includegraphics[scale=0.35,angle=0, bb = 550 20 240 600]{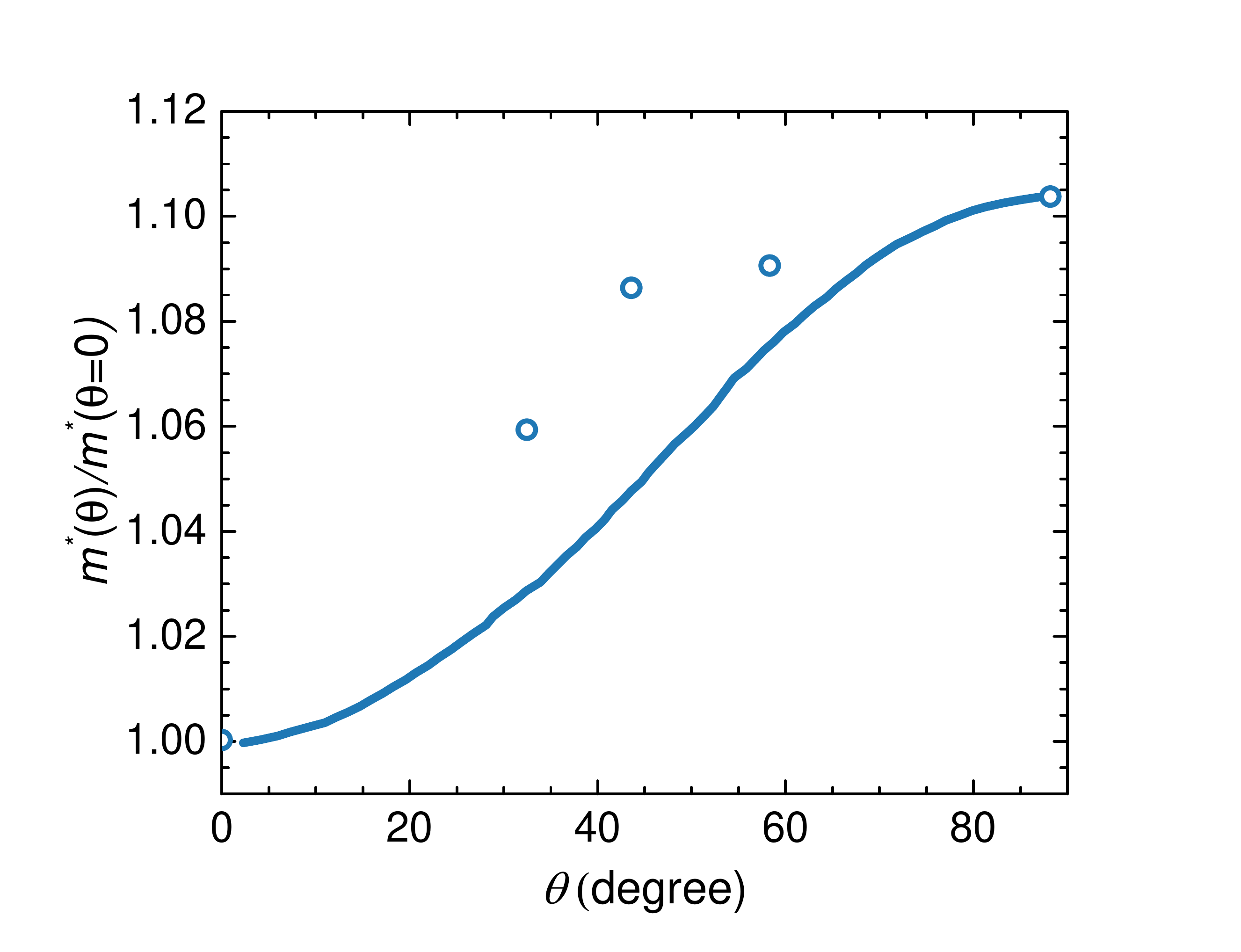}
\caption{\label{fig:epsart}{Normalized directional dependence of m$^*(\Theta)$ from SdH experiments [3] compared with theory for the electron concentration n = 2.36\,$ \times\,10^{18}$\,cm$^{-3}$.
}} \label{fig3th}
\end{figure}

By the least square method we have fitted simultaneously K as a function of $\Delta_{||}$ and m$^*(\Delta_{||})$ to the experimental data using five adjustable parameters: $P_{\perp}, P_{||}, {\cal E}_g, \delta, \Delta$ and obtained the following values: P$_{\perp} = (7.43 \pm 0.05)\times10^{-8}$ eVcm, P$_{||} = (7.21 \pm 0.05)\times10^{-8}$ eVcm, ${\cal E}_g = (-0.095 \pm 0.01)$ eV, $\delta = (0.085 \pm 0.01)$ eV, $\Delta = (0.27 \pm 0.03)$ eV. The negative value of ${\cal E}_g$ and positive value of $\delta$ indicate that Cd$_3$As$_2$ has the inverted band ordering with positive tetragonal field splitting. Theoretical dependence of K$^2$ and m${^*}^2$ upon 1/$\Delta_{||}$ (and the corresponding electron concentrations) obtained for our band model and the determined band parameters are compared with experimental data in Figs. 1 and 2 (curves 1). We show also theoretical predictions for simple level ordering with ${\cal E}_g$ = +0.025 eV (curve 2) and ${\cal E}_g$ = +0.14 eV (curve 3) with other parameters unchanged. It should be emphasized that a negative value of $\delta$ would give an increase of anisotropy ratio K$^2$ with concentration and thus it has been ruled out.

In Fig. 3 we plot the directional dependence of m$^*$. It is worth noting that the theoretical anisotropy of m$^*$ ($\approx$ 1.10) is considerably smaller than
the anisotropy of K$^2$ ($\approx$ 1.4) for the same sample, which is in agreement with Rosenman's data [2, 3]. For a simple generalization of Kane's model with anisotropy independent of the energy both quantities would have the same anisotropy.

\section{\label{sec:level2}Energy band model for C\MakeLowercase{d}$_3$A\MakeLowercase{s}$_2$}

Solutions of the secular equation (1) for our band parameters are shown in Fig. 4 for three directions of the wave vector \textbf{k}. The behavior of energy with \textbf{k} may suggest an overlap of the conduction and valence bands. However, the directional dependence of the energy with \textbf{k} clearly indicates that the conduction and the first valence band come into contact only at two \textbf{k} values and the energy gap becomes zero at these two points. Flat parts of the conduction and first valence band are a consequence of neglecting higher levels, which would have given a finite curvature of the bands, similar to those of HgTe under tensile stress [9].

\begin{figure}
\includegraphics[scale=0.3]{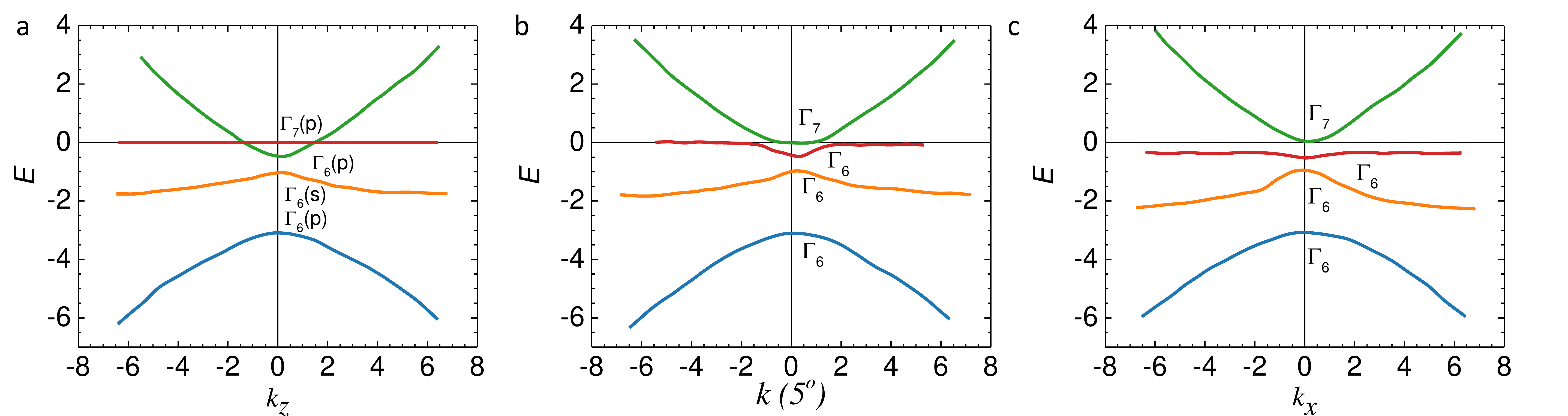}
\caption{\label{fig:epsart}{${\cal E}$ (\textbf{k}) relation for the four-level model of Cd$_3$As$_2$. The wave vector \textbf{k} (in $10^6$ cm$^{-1}$ units) is directed: a)  $0^{\circ}$, b)  $5^{\circ}$), c) $90^{\circ}$ from the tetragonal axis .The energy is in 10$^{-1}$ eV units.
}} \label{fig4th}
\end{figure}

The energy band model of Cd$_3$As$_2$ has been a controversial problem in the last years. Wagner et al [10] first suggested that Cd$_3$As$_2$ has the inverted level ordering. This was subsequently supported by Caron [11] and more recently by Cisowski and Arushanov [12] by  measurements of the thermoelectric power under pressure. As follows from the above analysis, the best fit to the SdH and dHvA data is obtained with the inverted band structure and positive crystal field splitting. Other models do not give good fit to the experiment with reasonable values of the band parameters.

\begin{acknowledgments}
I am pleased to thank Dr W. Zawadzki for his continuous help and interest in this work. I am also indebted to Prof. W. Zdanowicz, Dr J. Cisowski and Dr E. Arushanov for informative discussions.
\end{acknowledgments}

\end{document}